\documentclass[pdflatex,sn-mathphys]{sn-jnl}

\jyear{2021}%

\theoremstyle{thmstyleone}%
\theoremstyle{thmstyletwo}%

\theoremstyle{thmstylethree}%

\begin{document}

\title[Bismuth under Negative Pressure]{Could Negative Pressures Turn Bismuth into a Metal? The Case of the Expanded.}


\author[1]{\fnm{Flor B.} \sur{Quiroga}}
\equalcont{These authors contributed equally to this work.}

\author[1]{\fnm{David} \sur{Hinojosa-Romero}}
\equalcont{These authors contributed equally to this work.}

\author[2]{\fnm{Alexander} \sur{Valladares}}
\equalcont{These authors contributed equally to this work.}

\author[2]{\fnm{Renela M.} \sur{Valladares}}
\equalcont{These authors contributed equally to this work.}

\author[1]{\fnm{Isa\'{i}as} \sur{Rodr\'{i}guez}}
\equalcont{These authors contributed equally to this work.}

\author*[1]{\fnm{Ariel A.} \sur{Valladares}}\email{valladar@unam.mx}
\equalcont{These authors contributed equally to this work.}

\affil*[1]{\orgdiv{Instituto de Investigaciones en Materiales}, \orgname{Universidad Nacional Aut\'{o}noma de
M\'{e}xico}, \orgaddress{\street{Apartado Postal 70-360}, \city{Ciudad Universitaria}, \postcode{04510}, \state{CDMX}, \country{M\'{e}xico}}}

\affil[2]{\orgdiv{Facultad de Ciencias}, \orgname{Universidad Nacional Aut\'{o}noma de M\'{e}xico}, \orgaddress{\street{Apartado Postal 70-542}, \city{Ciudad Universitaria}, \postcode{04510}, \state{CDMX}, \country{M\'{e}xico}}}


\abstract{Materials may behave in non-expected ways when subject to unexpected
conditions. For example, when Bi was turned into an amorphous phase (\textit{a}-Bi)
unexpectedly it became a superconductor at temperatures below $10$ K. We provided an explanation as to why \textit{a}-Bi superconducts and the crystalline (\textit{c}-Bi) had not been found to do so: we computer calculated their electronic properties and found that \textit{a}-Bi has a larger electron density of states, eDoS, at the Fermi surface than \textit{c}-Bi and this explained the phenomenon. We even predicted an upper limit for the superconducting $T_c$ of the crystalline phase, which was experimentally corroborated within the following year. We now decided to investigate what happens to crystalline (Wyckoff structure) and amorphous Bi when pressures below the atmospheric are applied (expansion). Here we show that when expanded, \textit{c}-Bi becomes more metallic, since the eDoS increases when the volume increases for the Wyckoff structure, while the amorphous eDoS decreases. If the crystalline structure is maintained its $T_c$ would rise under expansion, whereas it would diminish for the \textit{a}-Bi. Expansion can be obtained in the laboratory by chemically etching Bi-based alloys, a process also known as dealloying, for example.}




\maketitle

\section{Introduction}\label{sec1}

The quest for a room temperature superconductor has permeated the field for decades. Patents and claims have come and gone and with the advent of laboratory equipment that produces very high pressures for very short periods of time the quest has increased. Recently, a (controversial) claim, portrayed in a paper on carbonaceous sulfur hydride \cite{Snider_2020} that supposedly is a room temperature superconductor under very high pressures, provoked an optimism that gave hope to accomplish this quest for more accessible pressures. However, the high pressures required made the original optimism thermalize the hopes and realize that since the difficulties involved in the process are considerable it does not seem possible to develop a room temperature superconductor at reasonable pressures. These experimental results were questioned by some researchers who did not give credit to the claim \cite{Hirsch_2021} (see reference \cite{Service_2021} also), disbelieve that has not been fully clarified. Faced with such controversy, we decided “to look the other way” and started asking what would happen if we consider pressures below atmospheric, “negative” pressures, and their effects on the electronic structure and vibrational properties of condensed matter.

Bismuth is an interesting material since it is one of the few elements that maintains some properties, like superconductivity. under varied circumstances. Bismuth was first found to be a superconductor when in the amorphous phase at ambient pressure \cite{Buckel_1954,Buckel_1956}. Then, when the Wyckoff crystalline phase (the structure at room temperature and atmospheric pressure) was subjected to positive pressures it changed crystalline structures but maintained the superconducting properties for most of the new topologies (See \cite{Rodriguez_2019} and \cite{Valladares_2018}, and references contained therein). Then we came into the play and predicted, using a simple BCS approach, that although the crystalline Wyckoff structure had not been found to superconduct, it should, for temperatures below $1.3$ mK \cite{Mata_Pinzon_2016}. A year after our prediction, experimentalists proved us correct and they found that the Wyckoff structure superconducts below $0.53$ mK at ambient pressure \cite{Prakash_2017}. We had ventured to estimate superconducting temperatures for bismuth under various conditions; for the crystalline phase, that had not been found to superconduct, we succeeded; our other predictions await verification \cite{Valladares_2018, Hinojosa_Romero_2018}. So, with this succinct background we decided that bismuth was the ideal “guinea pig” to test the effect of negative pressures and we undertook the search to see if it could become metallic or a semiconductor under expansion, and what the possibilities were to find it in a superconducting state. This is the origin of this work.

We shall report the changes that take place in the electronic structure of Wyckoff bismuth when expanded $5\%$, $10\%$, $15\%$ its ambient volume. We shall then consider the amorphous phases of these expanded structures to investigate the electronic changes that occur for both the crystal and the amorphous. To search for superconductivity, we have to calculate the vibrational densities of states and then apply the approach stated in Ref. \cite{Mata_Pinzon_2016}, which is not the subject of the present paper. For now, we shall be content to see what happens to the semi-metallicity of bismuth under expansion; whether it is maintained or whether it evolves to become a metal or a non-metal.

Studies on the modifications that occur under expansion of other solids have been reported in the literature like the work of Moruzzi and Marcus whom in 1989 investigated the electronic changes that occur is solid palladium under negative pressures  \cite{Moruzzi_1989}. They found that Pd may become magnetic when expanded.  Thirty years later, in 2019, we investigated the electronic properties of amorphous  \cite{Rodriguez_2019PRB} and amorphous-porous palladium (to be published). In both cases we found that when the stable crystalline structure of Pd is altered, magnetism appears. The field of expanding solids is not as well researched due to the difficulties of experimentally creating an expanded sample. However, developments, like the chemical etching of alloys, seem promising to foster an incursion in this field. Also, from the computational point of view this does not represent an insurmountable problem as we shall present in what follows.

In the following section we deal with the methodology that we used. Next, we present the results of our investigation, to go sequentially to analyze such results and infer possible consequences of our predictions. Finally, we elaborate on the conclusions as to what the implications are and to further our speculations for future work.

\section{Methodology}\label{sec2}
When dealing with materials that have not been experimentally realized one has to very careful not to create science fiction. In this work we assume that the crystalline Wyckoff structure, stable at room temperature and at atmospheric pressure, remains the same when expanded up to 15\%, which might not be the case. However, when amorphized, this handicap disappears and no memory of the original structure remains which then serves for comparison purposes.

\begin{figure}[h]
\centering
\includegraphics[width=0.9\textwidth]{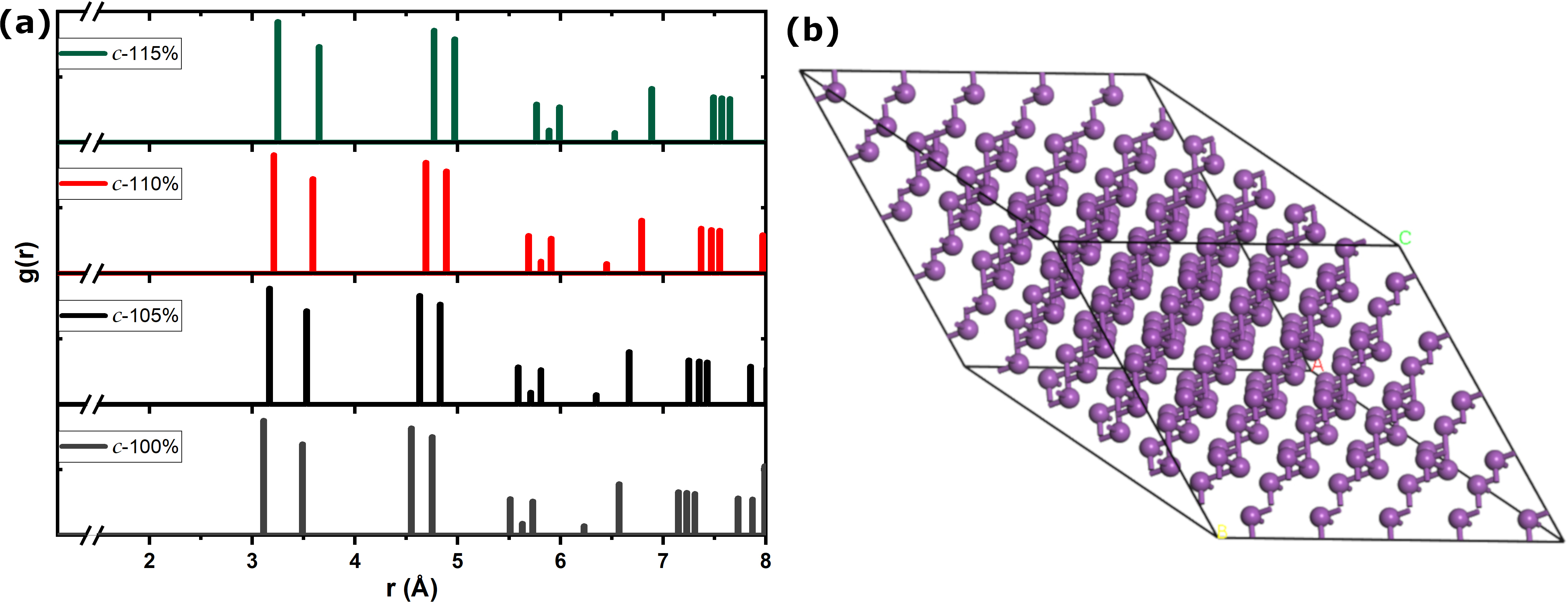}
\caption{(a) PDFs for the crystalline supercell structures. The positions of the peaks undergo a displacement consistent with the expansion forced on our structures. (b) 250-atom supercell of the Wyckoff structure where the bilayers can be seen. Linked atoms are nearest neighbors }\label{fig1}
\end{figure}

\begin{figure}[h]
\centering
\includegraphics[width=0.8\textwidth]{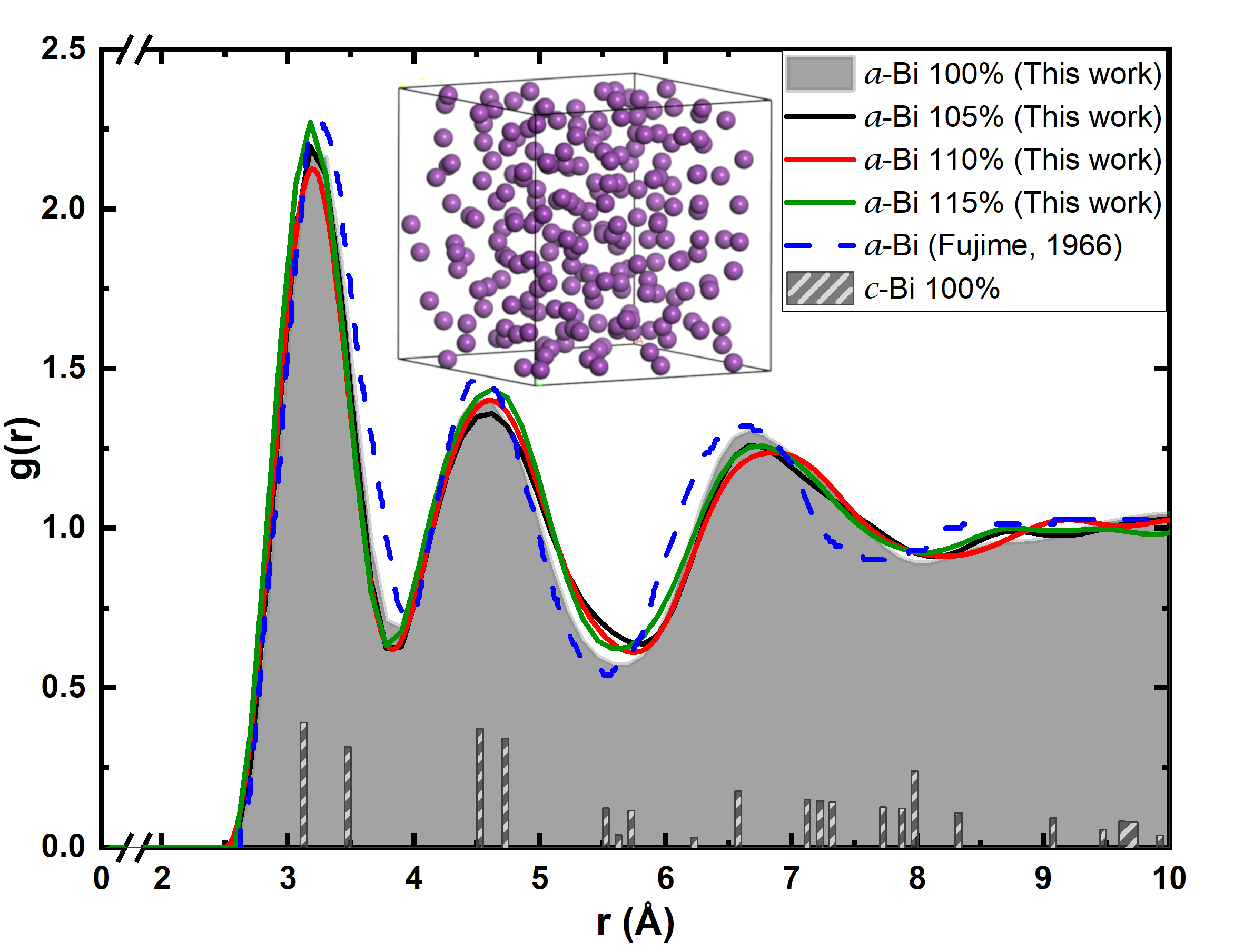}
\caption{PDFs for the amorphous supercells generated from the non-expanded and the expanded structures, after undergoing the MD \textit{undermelt-quench} procedure and the GO process. The experimental results reported by Fujime \citep{Fujime_1966} and the crystalline non-expanded PDFs, are also included for comparison. The 216-atom \textit{a}-Bi 100\% supercell is shown in the inset}\label{fig2}
\end{figure}

We constructed superlattices for the crystalline and for the amorphous samples.  For the crystalline structures the superlattices contained 250 atoms as a result of multiplying the Wyckoff rhombohedral unit cell $5\times5\times5$ times. For the amorphous we constructed a supercell with 216 atoms by multiplying the diamond-like structure $3\times3\times3$ times the cubic unit cell, an initially unstable structure to propitiate the evolution into a disordered topology, maintaining the correct density for the non-expanded structure ($9.81$ g/cm$^3$). After that we expanded both superlattices 5, 10 and 15\%, properly scaling the interatomic distances, to investigate the evolution of the electron densities of states. In Table \ref{tab1} we specify the lattice parameters and densities for all our samples. For the amorphous we ran our \textit{undermelt-quench} \citep{Valladares_2008} procedure to generate the corresponding samples. Figures \ref{fig1} and \ref{fig2} represent the Pair Distribution Functions (PDFs) for the crystalline and for the amorphous, respectively. In Figure \ref{fig2} we have included the experimental results by Fujime \citep{Fujime_1966}, and the agreement with our non-expanded amorphous sample is good. Also, in Figure \ref{fig2} we present the PDF for the non-expanded Wyckoff crystalline structure, for comparison.

\begin{table}[h]
\begin{center}
\begin{minipage}{174pt}
\caption{Parameters of our crystalline (\textit{c}-) and amorphous (\textit{a}-) samples.}\label{tab1}
\begin{tabular}{@{}ccc@{}}
\toprule
\multirow{2}{*}{System} & \multirow{2}{*}{\begin{tabular}[c]{@{}c@{}}Lattice parameter\\ {[}\AA{]}\end{tabular}} & \multirow{2}{*}{\begin{tabular}[c]{@{}c@{}}Density\\ {[}g/cm$^3${]}\end{tabular}} \\
 &    &   \\ \midrule
\textit{c}-100\% & 23.73 & \multirow{2}{*}{9.81}  \\
\textit{a}-100\% & 19.70 & \\ \midrule
\textit{c}-105\% & 24.12 & \multirow{2}{*}{9.34}  \\
\textit{a}-105\% & 20.02 & \\ \midrule
\textit{c}-110\% & 24.50 & \multirow{2}{*}{8.91}  \\
\textit{a}-110\% & 20.33 & \\ \midrule
\textit{c}-115\% & 24.86 & \multirow{2}{*}{8.53}  \\
\textit{a}-115\% & 20.64 & \\ 
\botrule
\end{tabular}
\end{minipage}
\end{center}
\end{table}

For the sake of completeness and self-containment we briefly reproduce the description of the \textit{undermelt-quench} procedure \citep{Valladares_2008}. We start from an unstable crystalline structure, with the correct density, to help the amorphizing procedure; usually for the metallic systems we have studied, we chose a diamond-like structure to initiate the process. Then we do Molecular Dynamics (MD) starting from 300 K linearly heating the sample to just below the melting temperature (\textit{undermelt}). Next we linearly cool the sample down to close to absolute zero (\textit{quench}). At the end of this MD process the structure is essentially disordered, but evidently unstable. To favor the final locally-stable structure we perform a Geometry Optimization (GO) procedure which generates amorphous structures that agree well with experimental results when available; that is why we consider the process adequate to predict structures that are presently not known, with their corresponding PDFs \citep{Rodriguez_2019PRB}. Once the GO process is completed we proceed to calculate the electron density of states (eDoS) for all the structures generated. It should be clear at this point that the amorphous supercell generated has periodic boundary conditions but we claim that it represents the material for distances comparable to the dimensions of the supercell. This periodicity is spurious but helps to calculate the physics of properties that do not extend beyond the size of the superlattice.

For all calculations, we employed a Density Functional approach as implemented in the DMol$^3$ code \citep{Delley_1995}, part of the Materials Studio suite \citep{biovia_materials_2016}. The electronic treatment consisted in a double numerical basis set with d-functions polarization (dnd) and a $6.0$ \AA\ real-space cutoff. We used the VWN functional \citep{VWN_1980} within the Local Density Approximation for the computation of the exchange-correlation energy and DFT Semi-core PseudoPotentials (dspp) \citep{Delley_2002} for the core treatment. All calculations were spin-unrestricted with SCF density convergence of $1\times10^{-6}$ using a thermal smearing of $0.136$ eV.

For the construction of the amorphous samples, the MD process uses a Nos\'{e}-Hoover NVT thermostat with a Nos\'{e} Q-ratio of $0.5$ and a time step of $19.4$ fs. The \textit{undermelt-quench} heating ramp started at 300 K and reached 534 K in 100 steps, followed by a cooling ramp of $2.34$ K per step down to $5.16$ K. For the GO procedure we sought the convergence by using the following thresholds: $2.7\times10^{-4}$ eV, $2\times10^{-3}$ eV \AA $^{-1}$ and $5\times10^{-3}$ \AA\ for energy, maximum force and maximum displacement, respectively.

The single-point energy and eDoS calculations for each geometrical structure were conducted at the $\Gamma$-point of the (periodic) amorphous and crystalline supercells, and also for a $2\times2\times2$ set of \textit{k}-points in the first Brillouin zone. In order to represent a bulk material, we applied a $0.2$ eV gaussian broadening to the eigenvalues obtained from each eDoS calculation.

A word on nomenclature, the non-expanded samples are referred to as the 100\% structures, whereas the samples expanded $x$\% are referred to as (10$x$)\%, or to (1$x$)\%, when $x$ is a two-digit number. 

\section{Results and Analysis}\label{sec3}
\begin{figure}[H]
\centering
\includegraphics[width=0.8\textwidth]{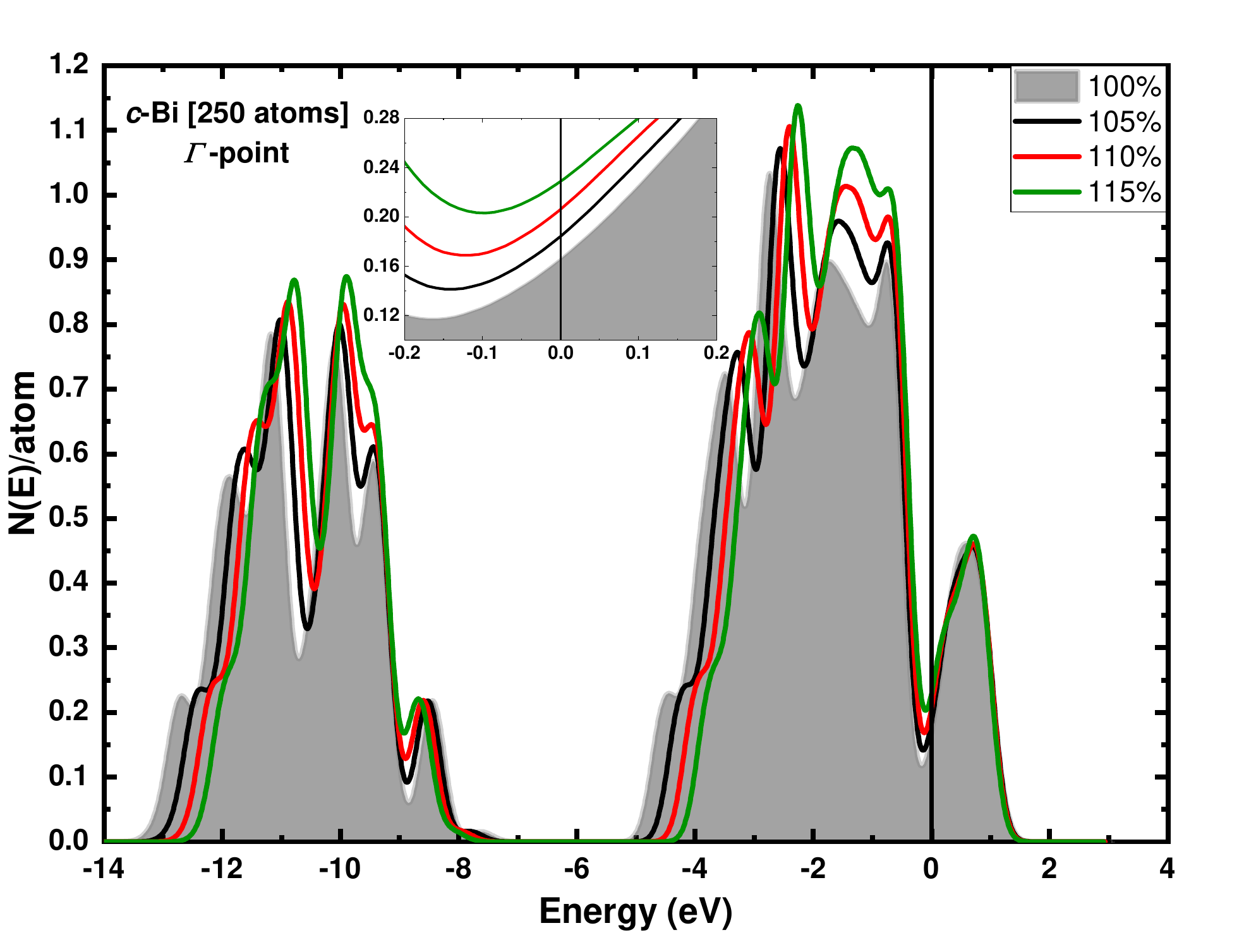}
\caption{eDoS per atom at the $\Gamma$-point for the crystalline non-expanded Wyckoff structure and for the expanded supercells. The inset shows the behavior of the curves in the vicinity of the Fermi energy}\label{fig3}
\end{figure}

\begin{figure}[H]
\centering
\includegraphics[width=0.8\textwidth]{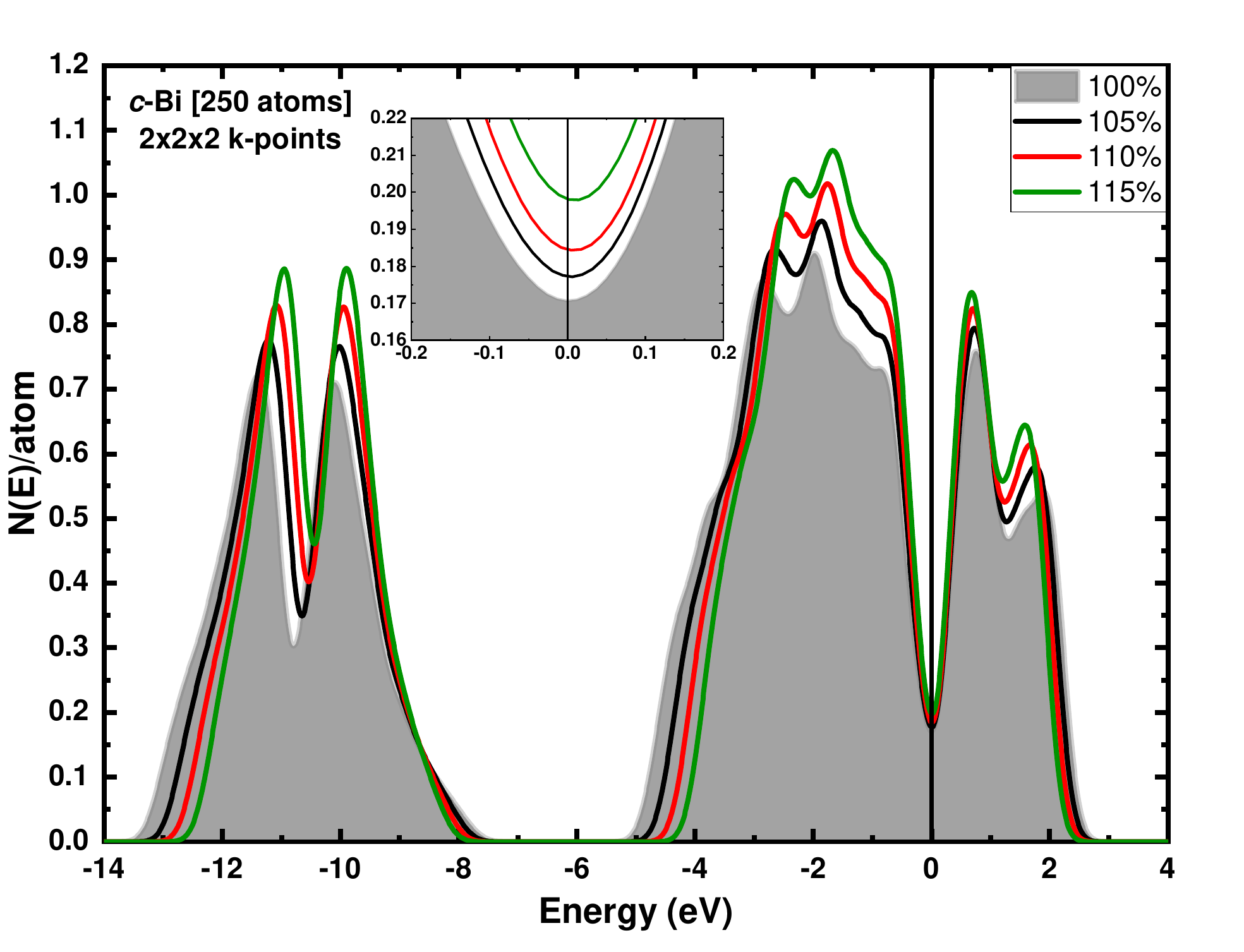}
\caption{eDoS per atom for $2\times2\times2$ \textit{k}-points in the first Brillouin zone for the crystalline non-expanded Wyckoff structure and for the expanded supercells. The inset shows the behavior of the curves in the vicinity of the Fermi energy}\label{fig4}
\end{figure}

\begin{figure}[H]
\centering
\includegraphics[width=0.8\textwidth]{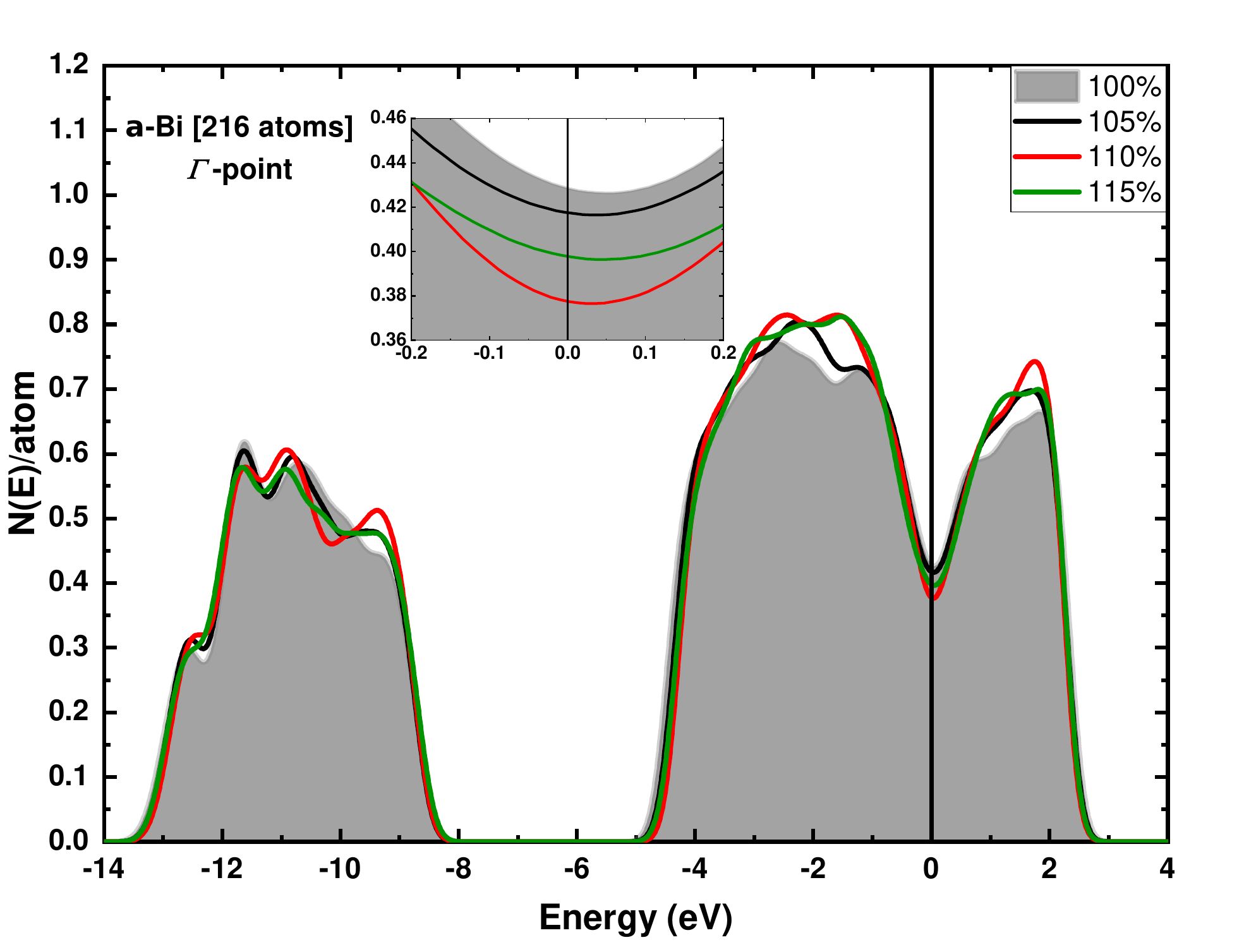}
\caption{eDoS per atom at the $\Gamma$-point for the amorphous expanded and non-expanded structures after the GO process. The inset shows the behavior of the curves in the vicinity of the Fermi energy}\label{fig5}
\end{figure}

\begin{figure}[H]
\centering
\includegraphics[width=0.8\textwidth]{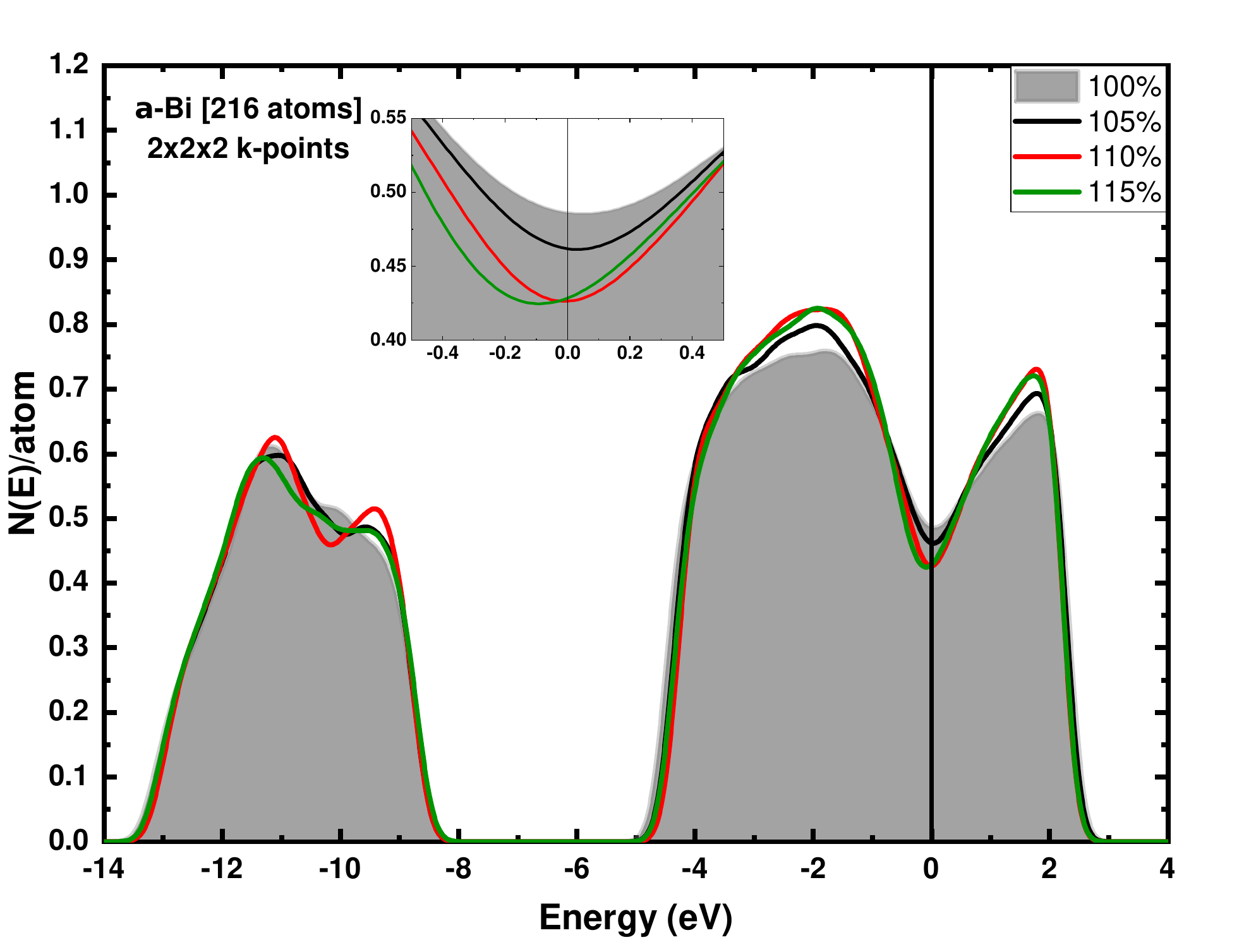}
\caption{eDoS per atom for $2\times2\times2$ \textit{k}-points in the first Brillouin zone for the amorphous expanded and non-expanded supercells after the GO process. The inset shows the behavior of the curves in the vicinity of the Fermi energy}\label{fig6}
\end{figure}

\begin{figure}[H]
\centering
\includegraphics[width=0.9\textwidth]{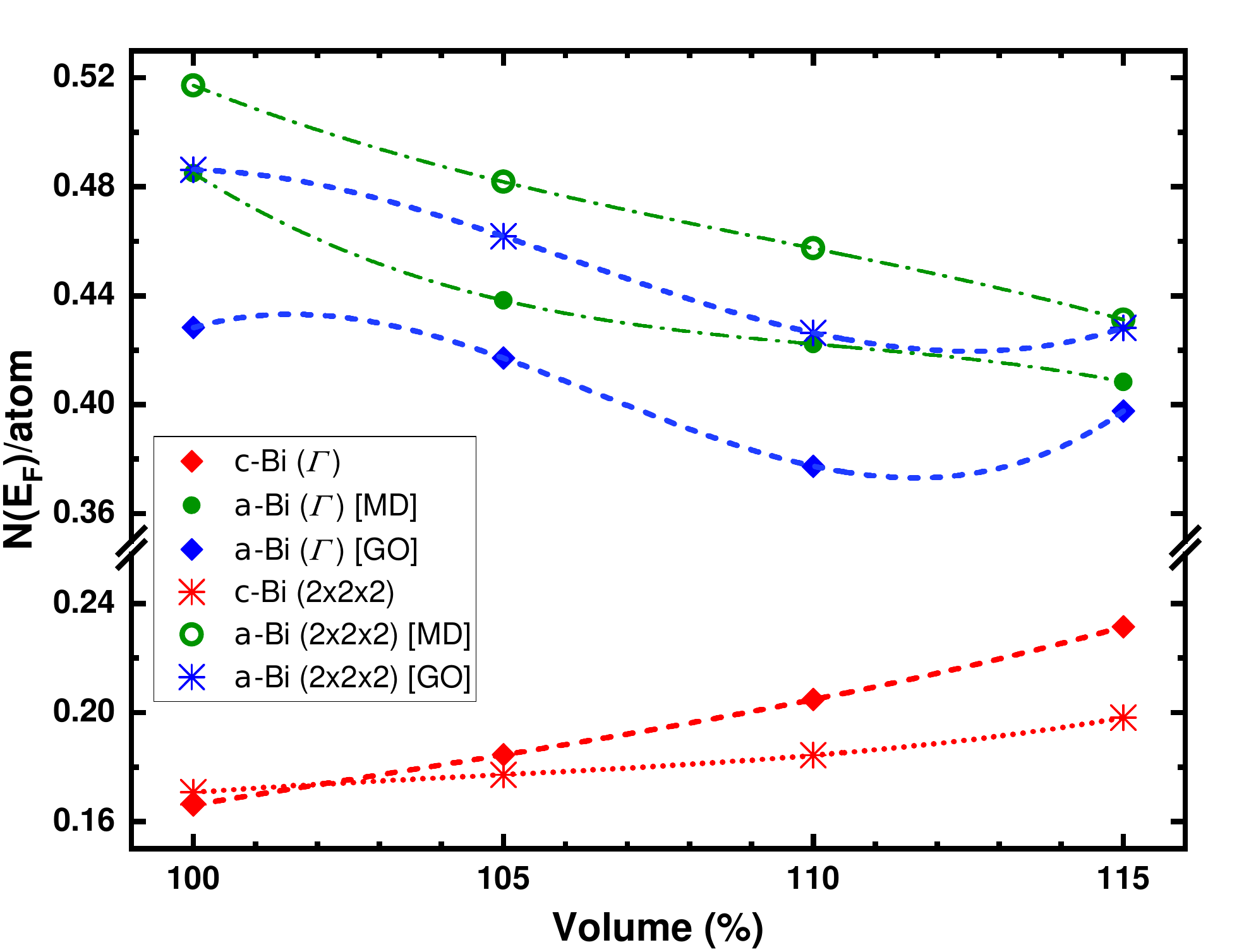}
\caption{Comparison of the eDoS at the Fermi level for all the samples studied after the MD and GO procedures. The results correspond to the $\Gamma$-point and to the $2\times2\times2$ set of \textit{k}-points in the first Brillouin zone of the crystalline and amorphous supercell. The curves are adjusted cubics to our results}\label{fig7}
\end{figure}

The calculated eDoS are represented in Figures \ref{fig3} and \ref{fig4} for the crystalline structures, and in Figures \ref{fig5} and \ref{fig6} for the amorphous samples after the GO processes. We also obtained the eDoS for the structures resulting from the MD processes, Figure \ref{fig7}; the reason to present these results for the non-optimized superlattices is to generate eDoS that may evolve, in the laboratory, towards the optimized ones in search for stable topologies and also to observe the changes that take place due to the optimization procedure. Since the runs for the crystalline are done on supercells with 250 atoms, whereas for the amorphous we considered supercells with 216 atoms, we report the eDoS per atom to normalize our results and to compare them readily.

Figure \ref{fig7} displays the results for the values of the eDoS at the Fermi level where trends that appear to be systematic are shown. No sudden changes occur anywhere for the expansions studied, contrary to what was observed in Ref. \citep{Hinojosa_Romero_2017}, which may indicate a tendency to possible modifications of the structure. However, a geometry optimization of the expanded crystalline samples is in order to elucidate this point.

Since our purpose is to investigate the behavior of the eDoS curves in the vicinity of the Fermi energy, in Fig. \ref{fig3} to \ref{fig6} we show this behavior in the insets for the structures considered. Since the parameters used for these calculations are the same for all samples, the results are directly contrasted. In this manner we can concentrate on visualizing, comparing and analyzing all our results, for both non-expanded and expanded, crystalline and amorphous, structures. An expected result is that the amorphous samples have a larger eDoS than the crystalline ones in all cases studied here. This was expected because that was the origin of our proposal to predict that the Wyckoff structure should superconduct below $1.3$ mK \citep{Mata_Pinzon_2016}. However, an unexpected result is to see that the eDoS for the amorphous diminishes with the expansion, while the opposite occurs for the crystalline ones. This could be due to the fact that as the amorphous expands, the tendency is to become more crystalline-like since the appearance of pores due to the increase in volume fosters the formation of small regions where the structure is not condition by the amorphicity and tends to crystallize. The opposite happens with the expanded crystalline cells as they rearrange to cope with this expansion. It may be possible that eventually, the tendencies will approach one another and become indistinguishable. Further calculations are needed to discern this supposition.

What would be the effect of this behavior in the superconductivity of the cells studied? If we think in terms of the BCS theory, the superconducting transition temperatures $T_c$ would depend not only on the eDoS but on the vibrational densities of states (vDoS) and on the strength of the Cooper pairing. According to the theory,
\begin{equation*}
    T_c = 1.13\ \Theta_D \exp\left( -\dfrac{1}{N(E_F)V} \right),
\end{equation*}
where $\Theta_D$ is the Debye temperature of the material, $N(E_F)$ is the eDoS at the Fermi level that we have obtained, and $V$ is Cooper's electron-electron attractive potential. So it would be necessary to get information concerning $\Theta_D$ (through the vDoS) and $V$ to reach reasonable conclusions.

\section{Conclusions}\label{sec4}
From looking at Figure \ref{fig7} we conclude that the crystalline structures upon expansion become more conducting whereas the amorphous ones become less conducting. Clearly, no changes in the structure of the crystalline cells upon expansion have been considered and since the Wyckoff structure is one that would tend to become cubic (look at the closeness of the two nearest-neighbor peaks in Fig. \ref{fig1}) it would be likely that any perturbation on the topology may induce rearrangements. However, the idea of this work was to investigate any changes that would occur upon expansion maintaining the geometry of the crystalline supercells unchanged to see what the effect of the pressure “looking the other way” (negative pressures) would have on the electronic properties, and we have accomplished that. There are many directions to go from here and hopefully some will be followed as an alternative to studying the effects that positive pressures have on the physical properties of solids. The developments due to chemically etching alloys is an active field at the experimental level. It is likely that as it progresses new alternatives will be found to consider the effect of negative pressures on solids, in general, and in particular on bismuth.

Superconductivity is an appealing phenomenon to be studied, but it requires to calculate the vibrational density of states and the pairing strength proposed by Cooper to reach conclusions. These calculations will show whether the expansion becomes a promising process to find new superconductors, with higher transition temperatures, for the betterment of mankind.

\backmatter

\bmhead{Acknowledgments}

F.B.Q. and D.H.-R. acknowledge Consejo Nacional de Ciencia y Tecnología (CONACyT) for supporting their graduate studies. I.R. thanks PAPIIT, DGAPA-UNAM, for his postdoctoral fellowship. A.A.V., R.M.V. and A.V. thank DGAPA-UNAM (PAPIIT) for continued financial support to carry out research projects under Grants No. IN104617 and IN116520. Mar\'{i}a Teresa V\'{a}zquez and Oralia Jim\'{e}nez provided the information requested. Alberto Lopez and Alejandro Pompa assisted with the technical support and maintenance of the computing unit at IIM-UNAM. Simulations were partially carried at the Computing Center of DGTIC-UNAM through project LANCAD-UNAM-DGTIC-131.

\section*{Declarations}

\begin{itemize}
\item Funding: Financial support was received from DGAPA-UNAM (PAPIIT) under projects No. IN104617 and IN116520.
\item Conflict of interest/Competing interests: The authors have no competing interests to declare that are relevant to the content of this article.
\item Ethics approval: Not applicable.
\item Consent to participate: Not applicable.
\item Consent for publication: The authors gave their consent for the publication of this manuscript.
\item Availability of data and materials: The datasets analysed during the current study are available from the corresponding author on reasonable request. 
\item Code availability: Materials Studio is a licensed software acquired from Dassault Syst\`{e}mes BIOVIA.
\item Authors' contributions: Ariel A. Valladares, Alexander Valladares and Renela M. Valladares conceived this research and designed it with the participation of Flor B. Quiroga, David Hinojosa-Romero and Isa\'{i}as Rodr\'{i}guez. All the simulations were done by Flor B. Quiroga and David Hinojosa-Romero. All authors discussed and analyzed the results. Ariel A. Valladares wrote the first draft and the other authors enriched the manuscript.
\end{itemize}

\bibliography{Bi_ex}


\begin{thebibliography}{19}
\ifx \bisbn   \undefined \def \bisbn  #1{ISBN #1}\fi
\ifx \binits  \undefined \def \binits#1{#1}\fi
\ifx \bauthor  \undefined \def \bauthor#1{#1}\fi
\ifx \batitle  \undefined \def \batitle#1{#1}\fi
\ifx \bjtitle  \undefined \def \bjtitle#1{#1}\fi
\ifx \bvolume  \undefined \def \bvolume#1{\textbf{#1}}\fi
\ifx \byear  \undefined \def \byear#1{#1}\fi
\ifx \bissue  \undefined \def \bissue#1{#1}\fi
\ifx \bfpage  \undefined \def \bfpage#1{#1}\fi
\ifx \blpage  \undefined \def \blpage #1{#1}\fi
\ifx \burl  \undefined \def \burl#1{\textsf{#1}}\fi
\ifx \doiurl  \undefined \def \doiurl#1{\url{https://doi.org/#1}}\fi
\ifx \betal  \undefined \def \betal{\textit{et al.}}\fi
\ifx \binstitute  \undefined \def \binstitute#1{#1}\fi
\ifx \binstitutionaled  \undefined \def \binstitutionaled#1{#1}\fi
\ifx \bctitle  \undefined \def \bctitle#1{#1}\fi
\ifx \beditor  \undefined \def \beditor#1{#1}\fi
\ifx \bpublisher  \undefined \def \bpublisher#1{#1}\fi
\ifx \bbtitle  \undefined \def \bbtitle#1{#1}\fi
\ifx \bedition  \undefined \def \bedition#1{#1}\fi
\ifx \bseriesno  \undefined \def \bseriesno#1{#1}\fi
\ifx \blocation  \undefined \def \blocation#1{#1}\fi
\ifx \bsertitle  \undefined \def \bsertitle#1{#1}\fi
\ifx \bsnm \undefined \def \bsnm#1{#1}\fi
\ifx \bsuffix \undefined \def \bsuffix#1{#1}\fi
\ifx \bparticle \undefined \def \bparticle#1{#1}\fi
\ifx \barticle \undefined \def \barticle#1{#1}\fi
\bibcommenthead
\ifx \bconfdate \undefined \def \bconfdate #1{#1}\fi
\ifx \botherref \undefined \def \botherref #1{#1}\fi
\ifx \url \undefined \def \url#1{\textsf{#1}}\fi
\ifx \bchapter \undefined \def \bchapter#1{#1}\fi
\ifx \bbook \undefined \def \bbook#1{#1}\fi
\ifx \bcomment \undefined \def \bcomment#1{#1}\fi
\ifx \oauthor \undefined \def \oauthor#1{#1}\fi
\ifx \citeauthoryear \undefined \def \citeauthoryear#1{#1}\fi
\ifx \endbibitem  \undefined \def \endbibitem {}\fi
\ifx \bconflocation  \undefined \def \bconflocation#1{#1}\fi
\ifx \arxivurl  \undefined \def \arxivurl#1{\textsf{#1}}\fi
\csname PreBibitemsHook\endcsname

\bibitem{Snider_2020}
\begin{barticle}
\bauthor{\bsnm{Snider}, \binits{E.}},
\bauthor{\bsnm{Dasenbrock-Gammon}, \binits{N.}},
\bauthor{\bsnm{McBride}, \binits{R.}},
\bauthor{\bsnm{Debessai}, \binits{M.}},
\bauthor{\bsnm{Vindana}, \binits{H.}},
\bauthor{\bsnm{Vencatasamy}, \binits{K.}},
\bauthor{\bsnm{Lawler}, \binits{K.V.}},
\bauthor{\bsnm{Salamat}, \binits{A.}},
\bauthor{\bsnm{Dias}, \binits{R.P.}}:
\batitle{Room-temperature superconductivity in a carbonaceous sulfur hydride}.
\bjtitle{Nature}
\bvolume{586}(\bissue{7829}),
\bfpage{373}--\blpage{377}
(\byear{2020}).
\doiurl{10.1038/s41586-020-2801-z}
\end{barticle}
\endbibitem

\bibitem{Hirsch_2021}
\begin{barticle}
\bauthor{\bsnm{Hirsch}, \binits{J.E.}},
\bauthor{\bsnm{Marsiglio}, \binits{F.}}:
\batitle{Unusual width of the superconducting transition in a hydride}.
\bjtitle{Nature}
\bvolume{596}(\bissue{7873}),
\bfpage{9}--\blpage{10}
(\byear{2021}).
\doiurl{10.1038/s41586-021-03595-z}
\end{barticle}
\endbibitem

\bibitem{Service_2021}
\begin{barticle}
\bauthor{\bsnm{Service}, \binits{R.F.}}:
\batitle{Superconductor finding draws pointed critique}.
\bjtitle{Science}
\bvolume{374}(\bissue{6567}),
\bfpage{520}--\blpage{521}
(\byear{2021}).
\doiurl{10.1126/science.acx9468}
\end{barticle}
\endbibitem

\bibitem{Buckel_1954}
\begin{barticle}
\bauthor{\bsnm{Buckel}, \binits{W.}},
\bauthor{\bsnm{Hilsch}, \binits{R.}}:
\batitle{Einflu{\ss} der {K}ondensation bei tiefen {T}emperaturen auf den
  elektrischen {W}iderstand und die {S}upraleitung f{\"{u}}r verschiedene
  {M}etalle}.
\bjtitle{Zeitschrift f\"{u}r Physik}
\bvolume{138}(\bissue{2}),
\bfpage{109}--\blpage{120}
(\byear{1954}).
\doiurl{10.1007/BF01337903}
\end{barticle}
\endbibitem

\bibitem{Buckel_1956}
\begin{barticle}
\bauthor{\bsnm{Buckel}, \binits{W.}},
\bauthor{\bsnm{Hilsch}, \binits{R.}}:
\batitle{Supraleitung und elektrischer {W}iderstand neuartiger
  {Z}inn-{W}ismut-{L}egierungen}.
\bjtitle{Zeitschrift f\"{u}r Physik}
\bvolume{146}(\bissue{1}),
\bfpage{27}--\blpage{38}
(\byear{1956}).
\doiurl{10.1007/BF01326000}
\end{barticle}
\endbibitem

\bibitem{Rodriguez_2019}
\begin{barticle}
\bauthor{\bsnm{Rodr{\'{\i}}guez}, \binits{I.}},
\bauthor{\bsnm{Hinojosa-Romero}, \binits{D.}},
\bauthor{\bsnm{Valladares}, \binits{A.}},
\bauthor{\bsnm{Valladares}, \binits{R.M.}},
\bauthor{\bsnm{{Valladares}}, \binits{A.A.}}:
\batitle{A facile approach to calculating superconducting transition
  temperatures in the bismuth solid phases}.
\bjtitle{Scientific Reports}
\bvolume{9}(\bissue{1}),
\bfpage{5256}
(\byear{2019}).
\doiurl{10.1038/s41598-019-41401-z}
\end{barticle}
\endbibitem

\bibitem{Valladares_2018}
\begin{barticle}
\bauthor{\bsnm{Valladares}, \binits{A.A.}},
\bauthor{\bsnm{Rodr{\'{\i}}guez}, \binits{I.}},
\bauthor{\bsnm{Hinojosa-Romero}, \binits{D.}},
\bauthor{\bsnm{Valladares}, \binits{A.}},
\bauthor{\bsnm{Valladares}, \binits{R.M.}}:
\batitle{Possible superconductivity in the {B}ismuth {IV} solid phase under
  pressure}.
\bjtitle{Scientific Reports}
\bvolume{8}(\bissue{1}),
\bfpage{5946}
(\byear{2018}).
\doiurl{10.1038/s41598-018-24150-3}
\end{barticle}
\endbibitem

\bibitem{Mata_Pinzon_2016}
\begin{barticle}
\bauthor{\bsnm{Mata-Pinz{\'{o}}n}, \binits{Z.}},
\bauthor{\bsnm{Valladares}, \binits{A.A.}},
\bauthor{\bsnm{Valladares}, \binits{R.M.}},
\bauthor{\bsnm{Valladares}, \binits{A.}}:
\batitle{Superconductivity in {B}ismuth. {A} {N}ew {L}ook at an {O}ld
  {P}roblem}.
\bjtitle{{PLOS} {ONE}}
\bvolume{11}(\bissue{1}),
\bfpage{0147645}
(\byear{2016}).
\doiurl{10.1371/journal.pone.0147645}
\end{barticle}
\endbibitem

\bibitem{Prakash_2017}
\begin{barticle}
\bauthor{\bsnm{Prakash}, \binits{O.}},
\bauthor{\bsnm{Kumar}, \binits{A.}},
\bauthor{\bsnm{Thamizhavel}, \binits{A.}},
\bauthor{\bsnm{Ramakrishnan}, \binits{S.}}:
\batitle{Evidence for bulk superconductivity in pure bismuth single crystals at
  ambient pressure}.
\bjtitle{Science}
\bvolume{355}(\bissue{6320}),
\bfpage{52}--\blpage{55}
(\byear{2017}).
\doiurl{10.1126/science.aaf8227}
\end{barticle}
\endbibitem

\bibitem{Hinojosa_Romero_2018}
\begin{barticle}
\bauthor{\bsnm{Hinojosa-Romero}, \binits{D.}},
\bauthor{\bsnm{Rodr\'{i}guez}, \binits{I.}},
\bauthor{\bsnm{Valladares}, \binits{A.}},
\bauthor{\bsnm{Valladares}, \binits{R.M.}},
\bauthor{\bsnm{Valladares}, \binits{A.A.}}:
\batitle{Possible superconductivity in {B}ismuth (111) bilayers. {T}heir
  electronic and vibrational properties from first principles}.
\bjtitle{{MRS} Advances}
\bvolume{3}(\bissue{6-7}),
\bfpage{313}--\blpage{319}
(\byear{2018}).
\doiurl{10.1557/adv.2018.119}
\end{barticle}
\endbibitem

\bibitem{Moruzzi_1989}
\begin{barticle}
\bauthor{\bsnm{Moruzzi}, \binits{V.L.}},
\bauthor{\bsnm{Marcus}, \binits{P.M.}}:
\batitle{Magnetism in fcc rhodium and palladium}.
\bjtitle{Physical Review B}
\bvolume{39}(\bissue{1}),
\bfpage{471}--\blpage{474}
(\byear{1989}).
\doiurl{10.1103/PhysRevB.39.471}
\end{barticle}
\endbibitem

\bibitem{Rodriguez_2019PRB}
\begin{barticle}
\bauthor{\bsnm{Rodr{\'{\i}}guez}, \binits{I.}},
\bauthor{\bsnm{Valladares}, \binits{R.M.}},
\bauthor{\bsnm{Hinojosa-Romero}, \binits{D.}},
\bauthor{\bsnm{Valladares}, \binits{A.}},
\bauthor{\bsnm{Valladares}, \binits{A.A.}}:
\batitle{Emergence of magnetism in bulk amorphous palladium}.
\bjtitle{Physical Review B}
\bvolume{100}(\bissue{2}),
\bfpage{024422}
(\byear{2019}).
\doiurl{10.1103/PhysRevB.100.024422}
\end{barticle}
\endbibitem

\bibitem{Valladares_2008}
\begin{bchapter}
\bauthor{\bsnm{Valladares}, \binits{A.A.}}:
\bctitle{A new approach to the ab initio generation of amorphous semiconducting
  structures. {E}lectronic and vibrational studies}.
In: \beditor{\bsnm{Wolf}, \binits{J.C.}},
\beditor{\bsnm{Lange}, \binits{L.}} (eds.)
\bbtitle{Glass Materials Research Progress},
pp. \bfpage{61}--\blpage{123}.
\bpublisher{Nova Science Publishers},
\blocation{New York}
(\byear{2008})
\end{bchapter}
\endbibitem

\bibitem{Fujime_1966}
\begin{barticle}
\bauthor{\bsnm{Fujime}, \binits{S.}}:
\batitle{Electron {D}iffraction at {L}ow {T}emperature {II}. {R}adial
  {D}istribution {A}nalysis of {M}etastable {S}tructure of {M}etal {F}ilms
  {P}repared by {L}ow {T}emperature {C}ondensation}.
\bjtitle{Japanese Journal of Applied Physics}
\bvolume{5}(\bissue{9}),
\bfpage{764}--\blpage{777}
(\byear{1966}).
\doiurl{10.1143/JJAP.5.764}
\end{barticle}
\endbibitem

\bibitem{Delley_1995}
\begin{bchapter}
\bauthor{\bsnm{Delley}, \binits{B.}}:
\bctitle{{D}{M}ol, a {S}tandard {T}ool for {D}ensity {F}unctional
  {C}alculations: {R}eview and {A}dvances}.
In: \beditor{\bsnm{Seminario}, \binits{J.M.}},
\beditor{\bsnm{Politzer}, \binits{P.}} (eds.)
\bbtitle{Modern {D}ensity {F}unctional {T}heory: {A} {Tool} For {C}hemistry}
vol. \bseriesno{2},
pp. \bfpage{221}--\blpage{254}.
\bpublisher{Elsevier Science B. V.},
\blocation{Amsterdam}
(\byear{1995})
\end{bchapter}
\endbibitem

\bibitem{biovia_materials_2016}
\begin{botherref}
\oauthor{\bsnm{{Dassault Syst\`{e}mes BIOVIA}}}:
Materials {Studio} 2016: {DMol3}, {Forcite}.
Dassault Syst\`{e}mes BIOVIA
(2016)
\end{botherref}
\endbibitem

\bibitem{VWN_1980}
\begin{barticle}
\bauthor{\bsnm{Vosko}, \binits{S.H.}},
\bauthor{\bsnm{Wilk}, \binits{L.}},
\bauthor{\bsnm{Nusair}, \binits{M.}}:
\batitle{Accurate spin-dependent electron liquid correlation energies for local
  spin density calculations: a critical analysis}.
\bjtitle{Canadian Journal of Physics}
\bvolume{58}(\bissue{8}),
\bfpage{1200}--\blpage{1211}
(\byear{1980}).
\doiurl{10.1139/p80-159}
\end{barticle}
\endbibitem

\bibitem{Delley_2002}
\begin{barticle}
\bauthor{\bsnm{Delley}, \binits{B.}}:
\batitle{Hardness conserving semilocal pseudopotentials}.
\bjtitle{Physical Review B}
\bvolume{66}(\bissue{15}),
\bfpage{155125}
(\byear{2002}).
\doiurl{10.1103/PhysRevB.66.155125}
\end{barticle}
\endbibitem

\bibitem{Hinojosa_Romero_2017}
\begin{barticle}
\bauthor{\bsnm{Hinojosa-Romero}, \binits{D.}},
\bauthor{\bsnm{Rodr{\'{\i}}guez}, \binits{I.}},
\bauthor{\bsnm{Mata-Pinz{\'{o}}n}, \binits{Z.}},
\bauthor{\bsnm{Valladares}, \binits{A.}},
\bauthor{\bsnm{Valladares}, \binits{R.}},
\bauthor{\bsnm{Valladares}, \binits{A.A.}}:
\batitle{Compressed {C}rystalline {B}ismuth and {S}uperconductivity
  {\textemdash} {A}n ab initio computational {S}imulation}.
\bjtitle{{MRS} Advances}
\bvolume{2}(\bissue{9}),
\bfpage{499}--\blpage{506}
(\byear{2017}).
\doiurl{10.1557/adv.2017.66}
\end{barticle}
\endbibitem

\end{thebibliography}


\end{document}